\documentclass{article}
\usepackage{spconf}
\usepackage[utf8]{inputenc} % allow utf-8 input
\usepackage[T1]{fontenc}    % use 8-bit T1 fonts
\usepackage{hyperref}       % hyperlinks
\usepackage{url}            % simple URL typesetting
\usepackage{booktabs}       % professional-quality tables
\usepackage{amsfonts}       % blackboard math symbols
\usepackage{nicefrac}       % compact symbols for 1/2, etc.
\usepackage{microtype}      % microtypography
\usepackage{graphicx}   % manages graphics in Latex
\usepackage{url}    % allows sensible printing of URLs
\usepackage{fancyhdr}   % allows headers and footers
\usepackage{lastpage}   % provides page number of last page
\usepackage[table]{xcolor}
\usepackage{indentfirst}
\usepackage{diagbox}
\usepackage{graphicx}
\usepackage{subcaption}
\usepackage{caption} 
\usepackage{amsmath}
\usepackage{amssymb}
\usepackage{amsthm}
\usepackage{thmtools,thm-restate}
\usepackage{paralist}
\usepackage{tabularx}
\usepackage{longtable}
\usepackage{multirow}
\usepackage{multicol}
\usepackage{fancyvrb}
\usepackage{framed}
\usepackage{comment}
\usepackage{algorithm}
\usepackage{algorithmic}
\usepackage{paralist}
\usepackage{listings}
\usepackage{multirow}
\usepackage{mathtools}
\usepackage{enumerate}
\usepackage{booktabs}
\usepackage{siunitx}
\usepackage{array}
\usepackage{times}
\usepackage{makecell}

\newcommand{\x}{\mathbf{x}}
\newcommand{\y}{\mathbf{y}}
\newcommand{\z}{\mathbf{z}}

\newcommand{\RR}{\mathbb{R}}

\newcommand{\defeq}{\vcentcolon=}

\DeclarePairedDelimiterX{\inp}[2]{\langle}{\rangle}{#1, #2}

\title{Convolutional-Recurrent Neural Networks for Speech Enhancement}
\name{Han Zhao$^\dagger$ \qquad Shuayb Zarar$^\star$ \qquad Ivan Tashev$^\star$ \qquad Chin-Hui Lee$^\ddagger$\thanks{The work was done when HZ was an intern at Microsoft Research.}}
\address{$^\dagger$ Machine Learning Department, Carnegie Mellon University, Pittsburgh, PA, USA \\
$^\star$ Microsoft Research, One Microsoft Way, Redmond, WA, USA \\
$^\ddagger$ School of Electrical and Computer Engineering, Georgia Institute of Technology, Atlanta, GA, USA}

\begin{document}
\maketitle
\begin{abstract}
We propose an end-to-end model based on convolutional and recurrent neural networks for speech enhancement. Our model is purely data-driven and does not make any assumptions about the type or the stationarity of the noise. In contrast to existing methods that use multilayer perceptrons (MLPs), we employ both convolutional and recurrent neural network architectures. Thus, our approach allows us to exploit local structures in both the frequency and temporal domains. By incorporating prior knowledge of speech signals into the design of model structures, we build a model that is more data-efficient and achieves better generalization on both seen and unseen noise. Based on experiments with synthetic data, we demonstrate that our model outperforms existing methods, improving PESQ by up to 0.6 on seen noise and 0.64 on unseen noise. 
\end{abstract}
\begin{keywords}
convolutional neural networks, recurrent neural networks, speech enhancement, regression model
\end{keywords}
\section{Introduction}
\label{sec:intro}
\noindent
Speech enhancement~\cite{tashev2009sound,loizou2013speech} is one of the corner stones of building robust automatic speech recognition (ASR) and communication systems. The problem is of especial importance nowadays where modern systems are often built using data-driven approaches based on large scale deep neural networks~\cite{hinton2012deep,amodei2016deep}. In this scenario, the mismatch between clean data used to train the system and the noisy data encountered when deploying the system will often degrade the recognition accuracy in practice, and speech enhancement algorithms work as a preprocessing module that help to reduce the noise in speech signals before they are fed into these systems. 

Speech enhancement is a classic problem that has attracted much research efforts for several decades in the community. By making assumptions on the nature of the underlying noise, statistical based approaches, including the spectral subtraction method~\cite{boll1979suppression}, the minimum mean-square error log-spectral method~\cite{ephraim1985speech}, etc., can often obtain analytic solutions for noise suppression. However, due to these unrealistic assumptions, most of these statistical-based approaches often fail to build estimators that can well approximate the complex scenarios in real-world. As a result, additional noisy artifacts are usually introduced in the recovered signals~\cite{hussain2007nonlinear}.

\textbf{Related Work}. Due to the availability of high-quality, large-scale data and the rapidly growing computational resources, data-driven approaches using regression-based deep neural networks have attracted much interests and demonstrated substantial performance improvements over traditional statistical-based methods~\cite{lu2013speech,xu2014experimental,xu2015regression,mirsamadi2016causal,qian2017speech}. The general idea of using deep neural networks, or more specifically, the MLPs for noise reduction is not new~\cite{tamura1989analysis,xie1994family}, and dates back at least to~\cite{tamura1988noise}. In these works, MLPs are applied as general nonlinear function approximators to approximate the mapping from noisy utterance to its clean version. A multivariate regression-based objective is then optimized using numeric methods to fit model parameters. To capture the temporal nature of speech signals, previous works also introduced recurrent neural networks (RNNs)~\cite{maas2012recurrent}, which removes the needs for the explicit choice of context window in MLPs. 

\textbf{Contributions}. We propose an end-to-end model based on convolutional and recurrent neural networks for speech enhancement, which we term as \textsc{EHNet}. \textsc{EHNet} is purely data-driven and does not make any assumptions about the underlying noise. It consists of three components: the convolutional component exploits the local patterns in the spectrogram in both frequency and temporal domains, followed by a bidirectional recurrent component to model the dynamic correlations between consecutive frames. The final component is a fully-connected layer that predicts the clean spectrograms. Compared with existing models such as MLPs and RNNs, due to the sparse nature of convolutional kernels, \textsc{EHNet} is much more data-efficient and computationally tractable. Furthermore, the bidirectional recurrent component allows \textsc{EHNet} to model the dynamic correlations between consecutive frames adaptively, and achieves better generalization on both seen and unseen noise. Empirically, we evaluate the effectiveness of \textsc{EHNet} and compare it with state-of-the-art methods on synthetic dataset, showing that \textsc{EHNet} achieves the best performance among all the competitors on all the 5 metrics. Specifically, our model leads up to a 0.6 improvement of PESQ measure~\cite{recommendation2001perceptual} on seen noise and 0.64 improvement on unseen noise. 

\begin{figure*}[htb]
\centering
	\includegraphics[width=0.8\textwidth]{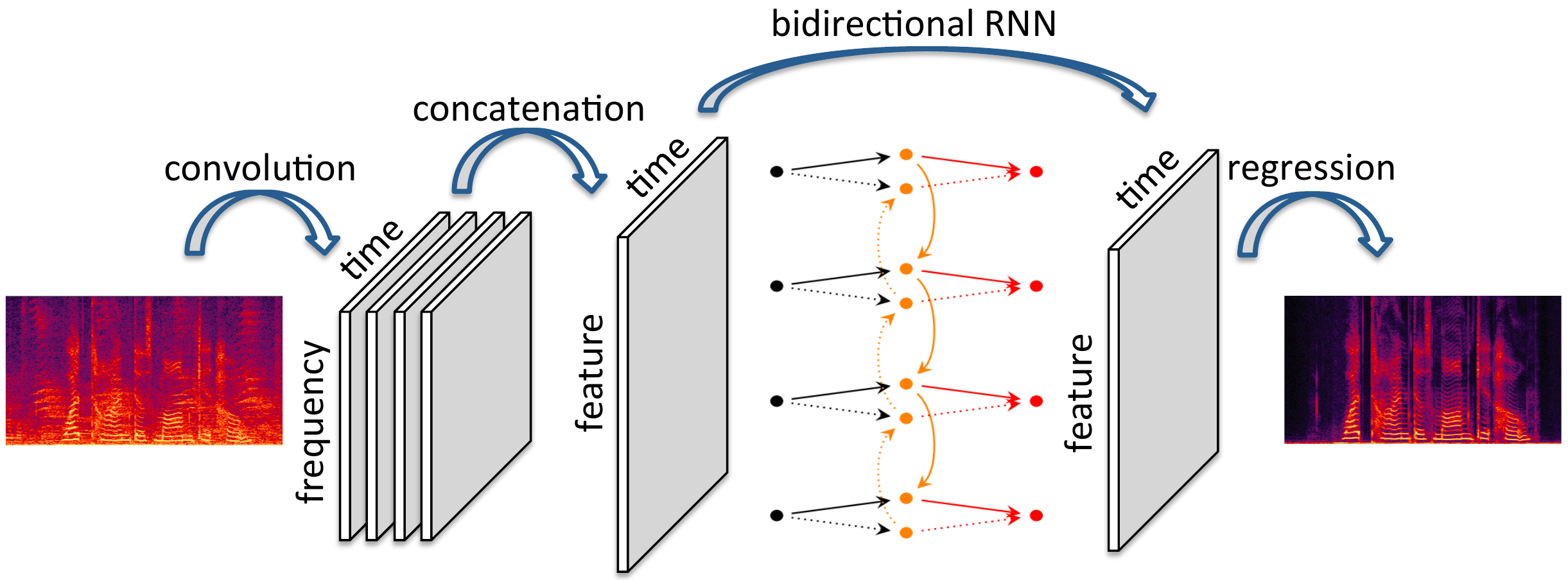}
\caption{Model architecture. \textsc{EHNet} consists of three components: noisy spectrogram is first convolved with kernels to form feature maps, which are then concatenated to form a 2D feature map. The 2D feature map is further transformed by a bidirectional RNN along the time dimension. The last component is a fully-connected network to predict the spectrogram frame-by-frame. \textsc{EHNet} can be trained end-to-end by defining a loss function between the predicted spectrogram and the clean spectrogram.}
\label{fig:model}
\end{figure*}

\section{Models and Learning}
\noindent
In this section we introduce the proposed model, \textsc{EHNet}, in detail and discuss its design principles as well as its inductive bias toward solving the enhancement problem. At a high level, we view the enhancement problem as a multivariate regression problem, where the nonlinear regression function is parametrized by the network in Fig.~\ref{fig:model}. Alternatively, the whole network can be interpreted as a complex filter for noise reduction in the frequency domain. 

\subsection{Problem Formulation}
\label{sec:formulation}
\noindent
Formally, let $\x\in \RR_+^{d\times t}$ be the noisy spectrogram and $\y\in\RR_+^{d\times t}$ be its corresponding clean version, where $d$ is the dimension of each frame, i.e., number of frequency bins in the spectrogram, and $t$ is the length of the spectrogram. Given a training set $\mathcal{D} = \{(\x_i, \y_i)\}_{i=1}^n$ of $n$ pairs of noisy and clean spectrograms, the problem of speech enhancement can be formalized as finding a mapping $g_\theta: \RR_+^{d\times t}\to\RR_+^{d\times t}$ that maps a noisy utterance to a clean one, where $g_\theta$ is parametrized by $\theta$. We then solve the following optimization problem to find the best model parameter $\theta$:
\begin{equation}
\min_\theta \quad \frac{1}{2}\sum_{i=1}^n ||g_\theta(\x_i) - \y_i||_F^2
\end{equation}
Under this setting, the key is to find a parametric family for denoising function $g_\theta$ such that it is both rich and data-efficient. 

\subsection{Convolutional Component}
\noindent
One choice for the denoising function $g_\theta$ is vanilla multilayer perceptrons, which has been extensively explored in the past few years~\cite{lu2013speech,xu2014experimental,xu2015regression,mirsamadi2016causal}. However, despite being universal function approximators~\cite{hornik1989multilayer}, the fully-connected network structure of MLPs usually cannot exploit the rich patterns existed in spectrograms. For example, as we can see in Fig.~\ref{fig:model}, signals in the spectrogram tend to be continuous along the time dimension, and they also have similar values in adjacent frequency bins. This key observation motivates us to apply convolutional neural networks to efficiently and cheaply extract local patterns from the input spectrogram. 

Let $\z\in\RR^{b\times w}$ be a convolutional kernel of size $b\times w$. We define a feature map $h_\z$ to be the convolution of the spectrogram $\x$ with kernel $\z$, followed by an elementwise nonlinear mapping $\sigma$: $h_\z(\x) = \sigma(\x * \z)$. Throughout the paper, we choose $\sigma(a) = \max\{a, 0\}$ to be the rectified linear function (ReLU), as it has been extensively verified to be effective in alleviating the notorious gradient vanishing problem in practice~\cite{maas2013rectifier}. Each such convolutional kernel $\z$ will produce a 2D feature map, and we apply $k$ separate convolutional kernels to the input spectrogram, leading to a collection of 2D feature maps $\{h_{\z_j}(\x)\}_{j=1}^k$. 

It is worth pointing out that without padding, with unit stride, the size of each feature map $h_\z(\x)$ is $(d-b+1)\times (t-w+1)$. However, in order to recover the original speech signal, we need to ensure that the final prediction of the model have exactly the same length in the time dimension as the input spectrogram. To this end, we choose $w$ to be an odd integer and apply a zero-padding of size $d\times \lfloor w/2\rfloor$ at both sides of $\x$ before convolution is applied to $\x$. This guarantees that the feature map $h_\z(\x)$ has $t + 2\times \lfloor w/2\rfloor - w + 1 = t + w - 1 - w + 1 = t$ time steps, matching that of $\x$.

On the other hand, because of the local similarity of the spectrogram in adjacent frequency bins, when convolving with the kernel $\z$, we propose to use a stride of size $b/2$ along the frequency dimension. As we will see in Sec.~\ref{sec:experiment}, such design will greatly reduce the number of parameters and the computation needed in the following recurrent component, without losing any prediction accuracy. 

\noindent\textbf{Remark}. We conclude this section by emphasizing that the application of convolution kernels is particularly well suited for speech enhancement in the frequency domain: each kernel can be understood as a nonlinear filter that detects a specific kind of local patterns existed in the noisy spectrograms, and the width of the kernel has a natural interpretation as the length of the context window. On the computational side, since convolution layer can also be understood as a special case of fully-connected layer with shared and sparse connection weights, the introduction of convolutions can thus greatly reduce the computation needed by a MLP with the same expressive power.

\subsection{Bidirectional Recurrent Component}
\noindent 
To automatically model the dynamic correlations between adjacent frames in the noisy spectrogram, we introduce bidirectional recurrent neural networks (BRNN) that have recurrent connections in both directions. The output of the convolutional component is a collection of $k$ feature maps $\{h_{\z_j}(\x)\}_{j=1}^k$, $h_{\z_j}(\x)\in\RR^{p\times t}$. Before feeding those feature maps into a BRNN, we need to first transform them into a 2D feature map:
$$H(\x) = [h_{\z_1}(\x); \ldots ; h_{\z_k}(\x)]\in \RR^{pk\times t}$$
In other words, we vertically concatenate $\{h_{\z_j}(\x)\}_{j=1}^k$ along the feature dimension to form a stacked 2D feature map $H(\x)$ that contains all the information from the previous convolutional feature map. 

In \textsc{EHNet}, we use deep bidirectional long short-term memory (LSTM)~\cite{hochreiter1997long} as our recurrent component due to its ability to model long-term interactions. At each time step $t$, given input $H_t\defeq H_t(\x)$, each unidirectional LSTM cell computes a hidden representation $\overrightarrow{H}_t$ using its internal gates:
\begin{align}
i_t &= s(W_{xi}H_t + W_{hi}\overrightarrow{H}_{t-1} + W_{ci}c_{t-1})\\
f_t &= s(W_{xf}H_t + W_{hf}\overrightarrow{H}_{t-1} + W_{cf}c_{t-1})\\
c_t &= f_t\odot c_{t-1} + i_t\odot\tanh(W_{xc}H_t + W_{hc}\overrightarrow{H}_{t-1})\\
o_t &= s(W_{xo}H_t + W_{ho}\overrightarrow{H}_{t-1} + W_{co}c_t) \\
\overrightarrow{H}_t &= o_t\odot\tanh(c_t)
\end{align}
where $s(\cdot)$ is the sigmoid function, $\odot$ means elementwise product, and $i_t$, $o_t$ and $f_t$ are the input gate, the output gate and the forget gate, respectively. The hidden representation $\tilde{H}_t$ of bidirectional LSTM is then a concatenation of both $\overrightarrow{H}_t$ and $\overleftarrow{H}_t$: $\tilde{H}_t = [\overrightarrow{H}_t; \overleftarrow{H}_t]$. To build deep bidirectional LSTMs, we can stack additional LSTM layers on top of each other. 

\subsection{Fully-connected Component and Optimization}
\noindent
Let $\tilde{H}(\x)\in \RR^{q\times t}$ be the output of the bidirectional LSTM layer. To obtain the estimated clean spectrogram, we apply a linear regression with truncation to ensure the prediction lies in the nonnegative orthant. Formally, for each $t$, we have:
\begin{equation}
\hat{\y}_t = \max\{0, W\tilde{H}_t + b_W\},\quad W\in\RR^{d\times q}, b_W\in\RR^d
\end{equation}
As discussed in Sec.~\ref{sec:formulation}, the last step is to define the mean-squared error between the predicted spectrogram $\hat{\y}$ and the clean one $\y$, and optimize all the model parameters simultaneously. Specifically, we use AdaDelta~\cite{zeiler2012adadelta} with scheduled learning rate~\cite{deng2013new} to ensure a stationary solution.

\section{Experiments}
\label{sec:experiment}
\noindent
To demonstrate the effectiveness of \textsc{EHNet} on speech enhancement, we created a synthetic dataset, which consists of 7,500, 1,500 and 1,500 recordings (clean/noisy speech) for training, validation and testing, respectively. Each recording is synthesized by convolving a randomly selected clean speech file with one of the 48 room impulse responses available and adding a randomly selected noise file. The clean speech corpus consists of 150 files containing ten utterances with male, female, and children voices. The noise dataset consists of 377 recordings representing 25 different types of noise. The room impulse responses were measured for distances between 1 and 3 meters. A secondary noise dataset of 32 files, with noises that do not appear in the training set, is denoted UnseenNoise and used to generate another test set of 1,500 files. The randomly generated speech and noise levels provide signal-to-noise ratio between 0 and 30 dB. All files are sampled with 16 kHz sampling rate and stored with 24 bits resolution.

\begin{table*}[htb]
\centering
\caption{Experimental results on synthetic dataset with both seen and unseen noise, evaluated with 5 different metrics. Noisy Speech corresponds to the scores obtained without enhancement, while Clean Speech corresponds to the scores obtained using the ground truth clean speech. For each metric, the model achieves the best performance is highlighted in bold.}
\label{table:result}
\begin{tabular}{|l||*5l||*5l|}\hline
 & \multicolumn{5}{c||}{\textbf{Seen Noise}} & \multicolumn{5}{c|}{\textbf{Unseen Noise}}\\\hline
\textbf{Model} & \textbf{SNR} & \textbf{LSD} & \textbf{MSE} & \textbf{WER} & \textbf{PESQ} & \textbf{SNR} & \textbf{LSD} & \textbf{MSE} & \textbf{WER} & \textbf{PESQ}\\\hline
Noisy Speech & 15.18 & 23.07 & 0.04399 & 15.40  & 2.26 & 14.78 & 23.76 & 	0.04786 & 18.4 & 2.09\\\hline
\textsc{MS} & 18.82 & 22.24 & 0.03985 & 14.77 & 2.40 & 19.73 & 22.82 & 0.04201 & \textbf{15.54}	& 2.26\\\hline
\textsc{DNN-Symm} & 44.51 & 19.89 & 0.03436 & 55.38 & 2.20 & 40.47 & 21.07 & 0.03741 & 54.77 & 2.16\\
\textsc{DNN-Causal} & 40.70 & 20.09 & 0.03485 & 54.92	 & 2.17 & 38.70 & 21.38 & 0.03718 & 54.13 & 2.13\\
\textsc{RNN-Ng} & 41.08 & 17.49 & 0.03533 & 44.93 & 2.19 & \textbf{44.60} & 18.81 & \textbf{0.03665} & 52.05 & 2.06\\\hline
\textsc{EHNet} & \textbf{49.79} & \textbf{15.17} & \textbf{0.03399} & \textbf{14.64} & \textbf{2.86} & 39.70 & \textbf{17.06} & 0.04712 & 16.71 & \textbf{2.73}\\\hline
Clean Speech & 57.31 & 1.01	& 0.00000 & 2.19 & 4.48 & 58.35 & 1.15 & 0.00000 & 1.83 & 4.48\\\hline
\end{tabular}
\end{table*}

\begin{figure*}[htb]
\centering
\begin{subfigure}[b]{0.15\textwidth}
	\includegraphics[width=\linewidth]{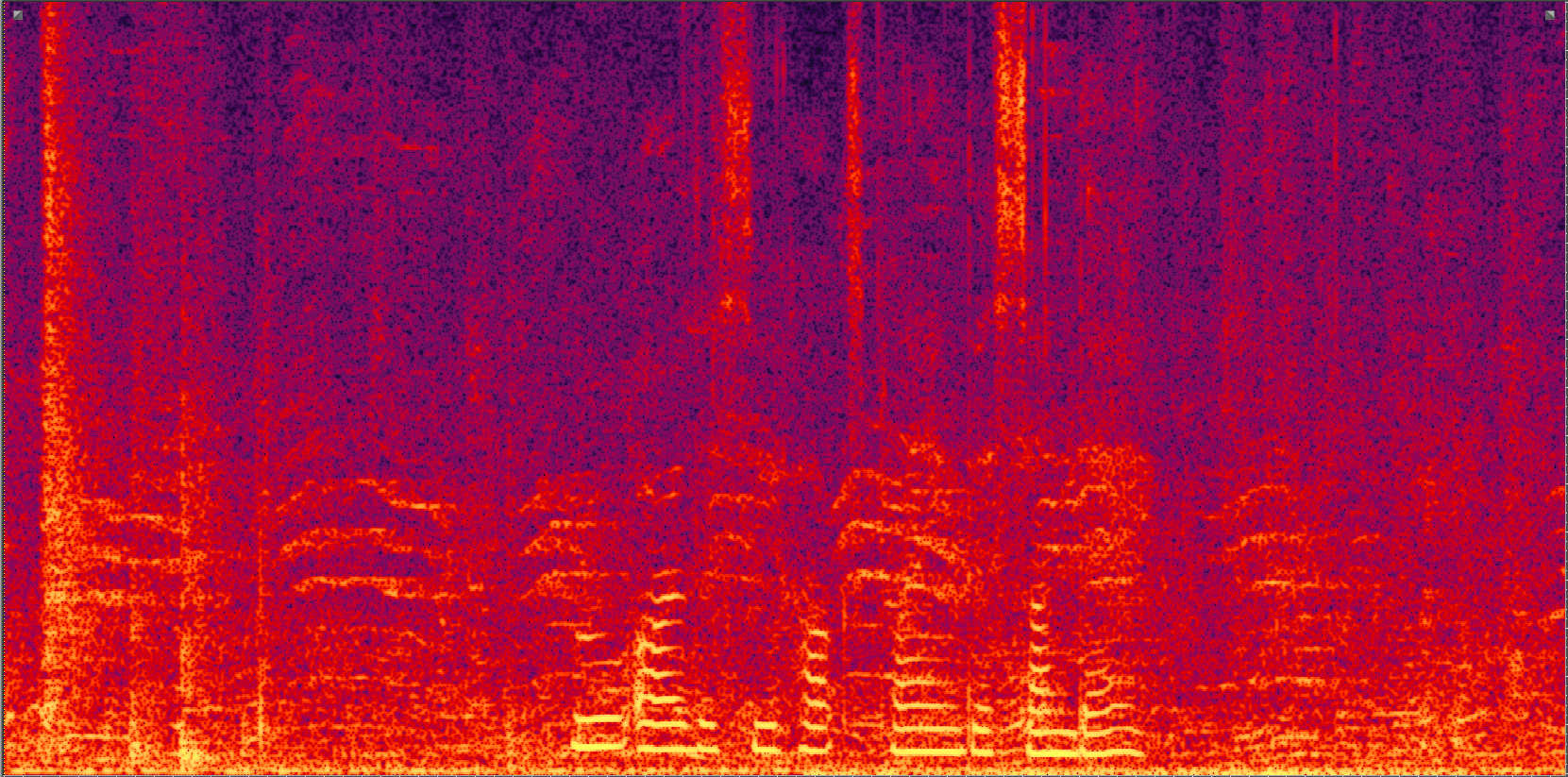}
	\caption{Noisy speech.}
\end{subfigure}
~
\begin{subfigure}[b]{0.15\textwidth}
	\includegraphics[width=\linewidth]{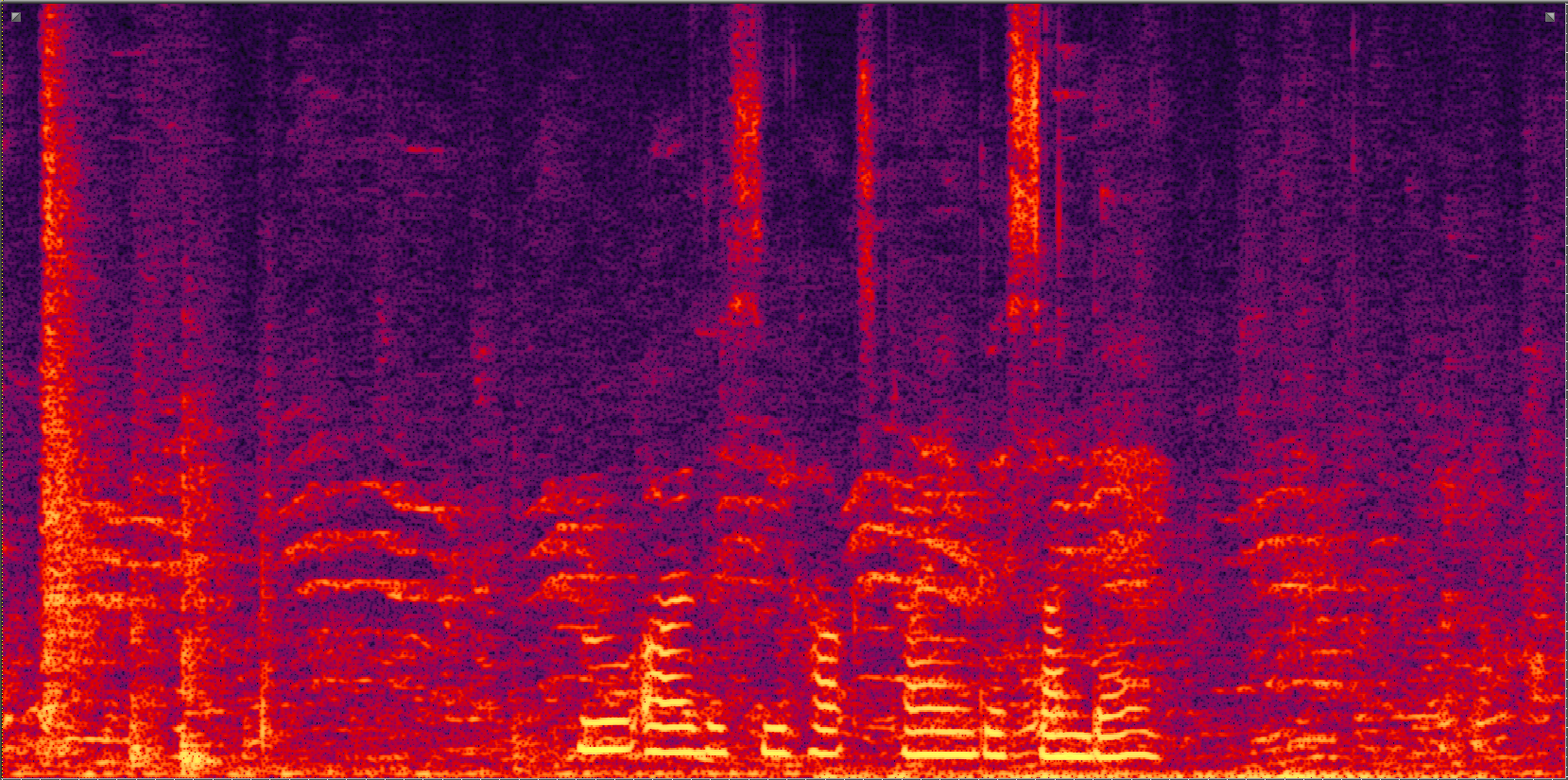}
	\caption{\textsc{MS}.}
\end{subfigure}
~
\begin{subfigure}[b]{0.15\textwidth}
	\includegraphics[width=\linewidth]{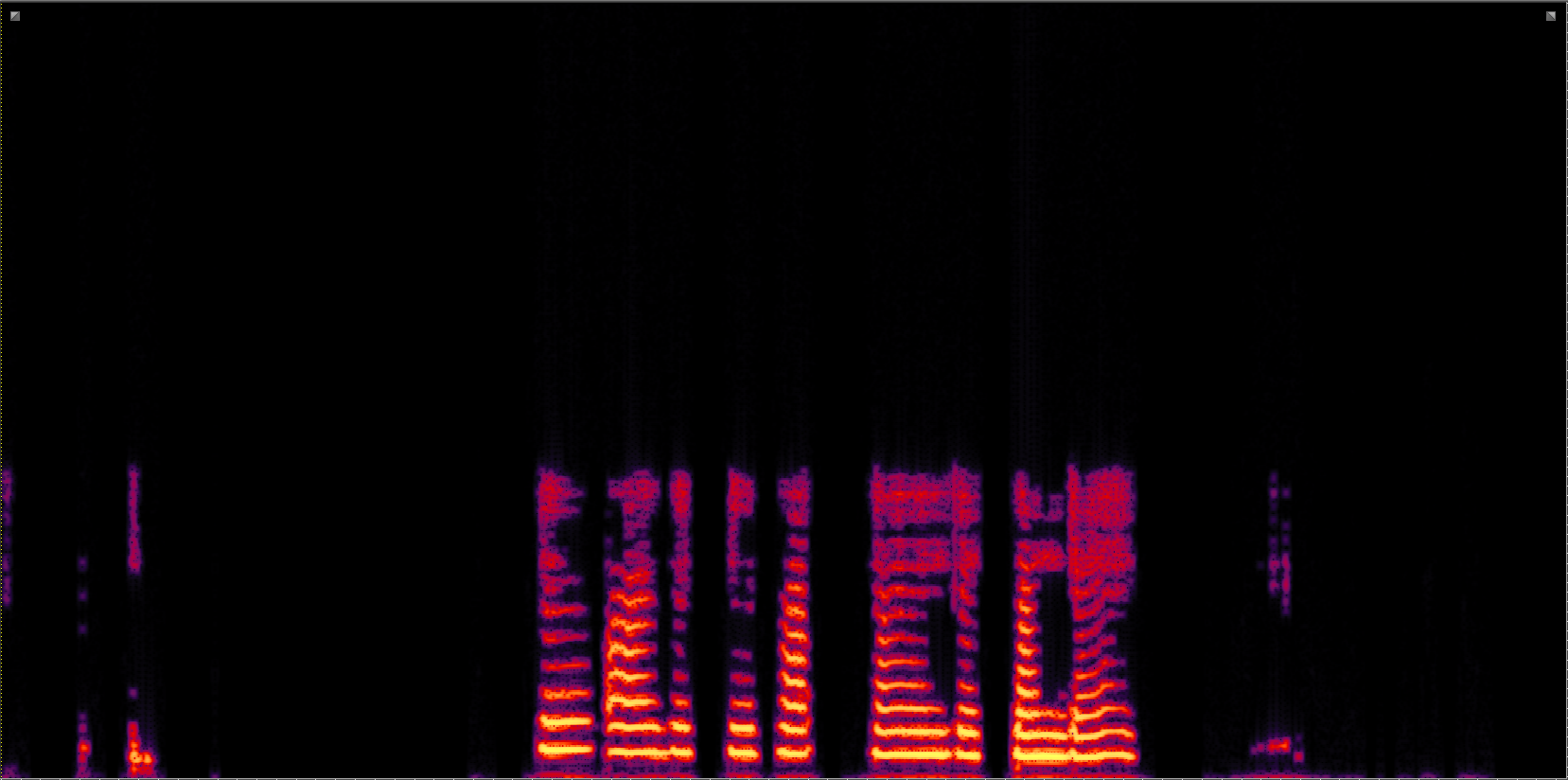}
	\caption{\textsc{DNN}.}
\end{subfigure}
~
\begin{subfigure}[b]{0.15\textwidth}
	\includegraphics[width=\linewidth]{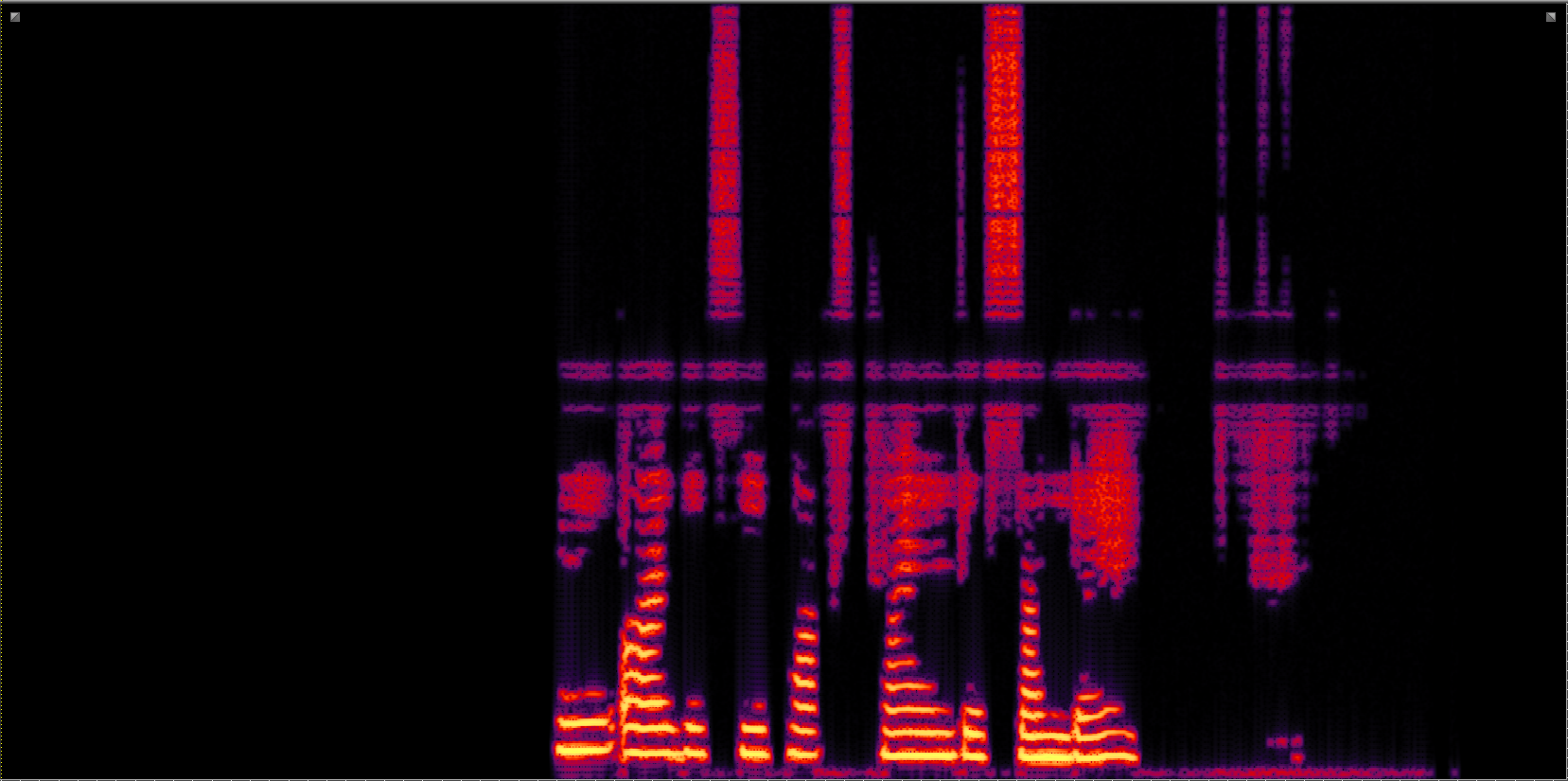}
	\caption{\textsc{RNN}.}
\end{subfigure}
~
\begin{subfigure}[b]{0.15\textwidth}
	\includegraphics[width=\linewidth]{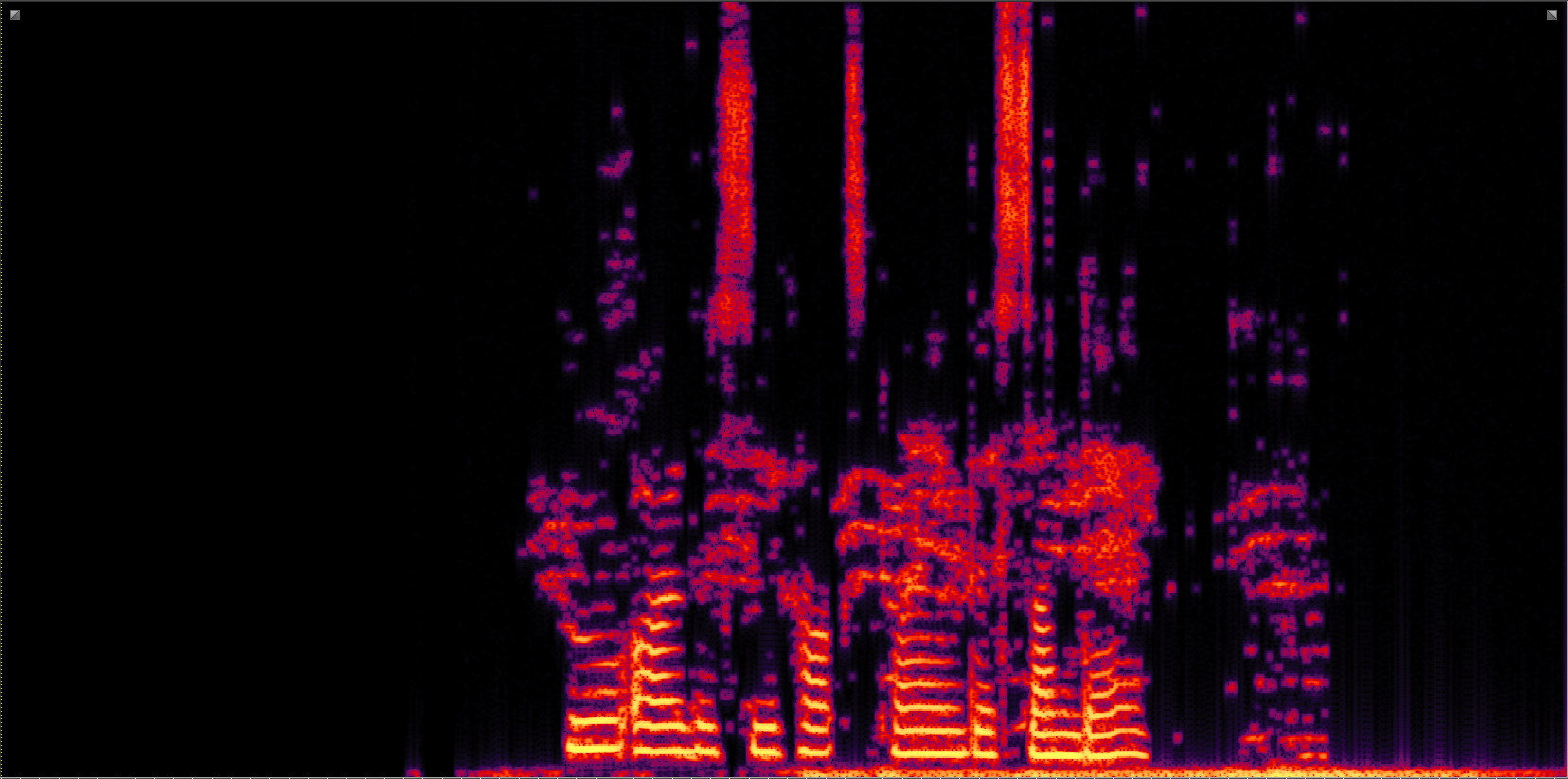}
	\caption{\textsc{EHNet}.}
\end{subfigure}
~
\begin{subfigure}[b]{0.15\textwidth}
	\includegraphics[width=\linewidth]{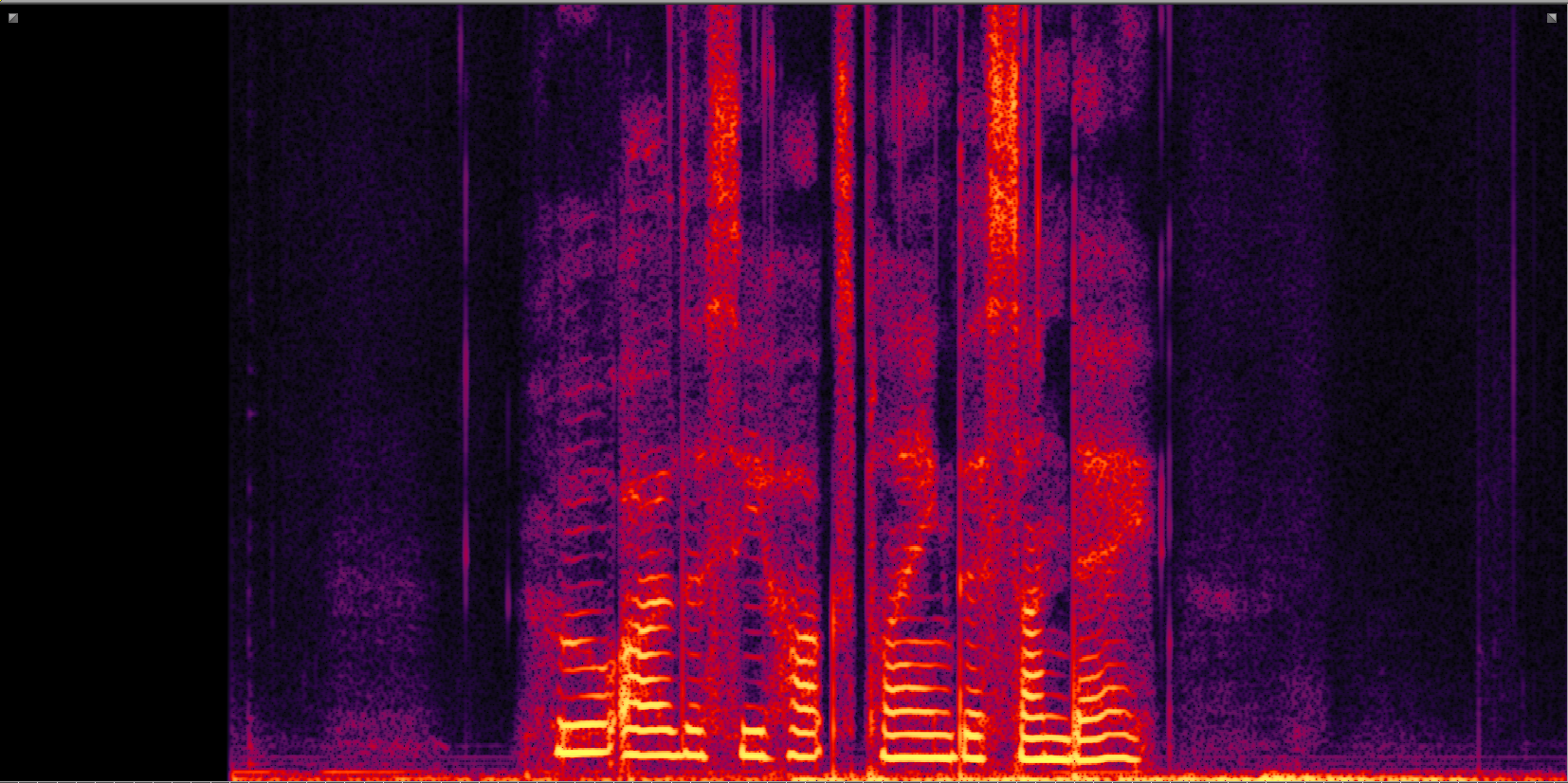}
	\caption{Clean speech.}
\end{subfigure}
\caption{Noisy and clean spectrograms, along with the denoised spectrograms using different models.}
\label{fig:case}
\end{figure*}

\subsection{Dataset and Setup}
\noindent
As a preprocessing step, we first use STFT to extract the spectrogram from each utterance. The spectrogram has 256 frequency bins $(d = 256)$ and $\sim 500$ frames $(t\approx 500)$ frames. To throughly measure the enhancement quality, we use the following 5 metrics to evaluate different models: signal-to-noise ratio (SNR, dB), log-spectral distortion (LSD), mean-squared-error on time domain (MSE), word error rate (WER, $\%$), and the PESQ measure. To measure WER, we use the DNN-based speech recognizer, described in~\cite{seide2011conversational}. The system is kept fixed (not fine-tuned) during the experiment. We compare our \textsc{EHNet} with the following state-of-the-art methods: 
\begin{enumerate}
	\item	\textsc{MS}. Microsoft's internal speech enhancement system used in production, which uses a combination of statistical-based enhancement rules. 
	\item 	\textsc{DNN-Symm}~\cite{xu2014experimental}. \textsc{DNN-Symm} contains 3 hidden layers, all of which have 2048 hidden units. It uses a symmetric context window of size 11. 
	\item 	\textsc{DNN-Causal}~\cite{mirsamadi2016causal}. Similar to \textsc{DNN-Symm}, \textsc{DNN-Causal} contains 3 hidden layers of size 2048, but instead of symmetric context window, it uses causal context window of size 7.
	\item 	\textsc{RNN-Ng}~\cite{maas2012recurrent}. \textsc{RNN-Ng} is a recurrent neural network with 3 hidden layers of size 500. The input at each time step covers frames in a context window of length 3. 
\end{enumerate}
The architecture of \textsc{EHNet} is as follows: the convolutional component contains 256 kernels of size $32\times 11$, with stride $16\times 1$ along the frequency and the time dimensions, respectively. We use two layers of bidirectional LSTMs following the convolution component, each of which has 1024 hidden units. To train \textsc{EHNet}, we fix the number of epochs to be 200, with a scheduled learning rate $\{1.0, 0.1, 0.01\}$ for every 60 epochs. For all the methods, we use the validation set to do early stopping and save the best model on validation set for evaluation on the test set. \textsc{EHNet} does not overfit, as both weight decay and dropout hurt the final performance. We also experiment with deeper \textsc{EHNet} with more layers of bidirectional LSTMs, but this does not significantly improve the final performance. We also observe in our experiments that reducing the stride of convolution in the frequency dimension does not significantly boost the performance of \textsc{EHNet}, but greatly incurs additional computations.

\subsection{Results and Analysis}
\noindent
Experimental results on the dataset is shown in Table~\ref{table:result}. On the test dataset with seen noise, \textsc{EHNet} consistently outperforms all the competitors with a large margin. Specifically, \textsc{EHNet} is able to improve the perceptual quality (PESQ measure) by 0.6 without hurting the recognition accuracy. This is very surprising as we treat the underlying ASR system as a black box and do not fine-tune it during the experiment. As a comparison, while all the other methods can boost the SNR ratio, they often decrease the recognition accuracy. More surprisingly, \textsc{EHNet} also generalizes to unseen noise as well, and it even achieves a larger boost (0.64) on the perceptual quality while at the same time increases the recognition accuracy. 

To have a better understanding on the experimental result, we do a case study by visualizing the denoised spectrograms from different models. As shown in Fig.~\ref{fig:case}, \textsc{MS} is the most conservative algorithm among all. By not removing much noise, it also keeps most of the real signals in the speech. On the other hand, although \textsc{DNN}-based approaches do a good job in removing the background noise, they also tend to remove the real speech signals from the spectrogram. This explains the reason why \textsc{DNN}-based approaches degrade the recognition accuracies in Table~\ref{table:result}. \textsc{RNN} does a better job than \textsc{DNN}, but also fails to keep the real signals in low frequency bins. As a comparison, \textsc{EHNet} finds a good tradeoff between removing background noise and preserving the real speech signals: it is better than \textsc{DNN}/\textsc{RNN} in preserving high/low-frequency bins and it is superior than \textsc{MS} in removing background noise. It is also easy to see that \textsc{EHNet} produces denoised spectrogram that is most close to the ground-truth clean spectrogram. 

\section{Conclusion}
\noindent
We propose \textsc{EHNet}, which combines both convolutional and recurrent neural networks for speech enhancement. The inductive bias of \textsc{EHNet} makes it well-suited to solve speech enhancement: the convolution kernels can efficiently detect local patterns in spectrograms and the bidirectional recurrent connections can automatically model the dynamic correlations between adjacent frames. Due to the sparse nature of convolutions, \textsc{EHNet} requires less computations than both MLPs and RNNs. Experimental results show that \textsc{EHNet} consistently outperforms all the competitors on all 5 different metrics, and is also able to generalize to unseen noises, confirming the effectiveness of \textsc{EHNet} in speech enhancement.

\newpage
\bibliographystyle{IEEEbib}
\bibliography{reference}

\end{document}